\documentclass{svjour3}                     % onecolumn (standard format)
\smartqed  % flush right qed marks, e.g. at end of proof
\usepackage{graphicx}
\newcommand\la{\langle}
\newcommand\ra{\rangle}

\newcommand\be{\begin{equation}}
\newcommand\ee{\end{equation}}
\newcommand\bea{\begin{eqnarray}}
\newcommand\eea{\end{eqnarray}}
\journalname{Few-Body Systems (FB20)}
\begin{document}

\title{
%Insert your title here
Relativistic Covariance of Light-Front Few-Body Systems in Hadron Physics
\thanks{Presented at the 20th International IUPAP Conference on Few-Body Problems in Physics, 20 - 25 August, 2012, Fukuoka, Japan}
}
%\subtitle{
%Do you have a subtitle?\\ If so, write it here
%Subtitle
%}


\author{Ho-Meoyng Choi         \and
        Chueng-Ryong Ji %etc.
}


\institute{H.-M. Choi \at
              Department of Physics Education, Kyungpook National University, Daegu 702-701, Korea  \\
%              Tel.: +123-45-678910\\
%              Fax: +123-45-678910\\
              \email{homyoung@knu.ac.kr}         %  \\
           \and
           C.-R. Ji \at
              Department of Physics, Box 8202, North Carolina State University, Raleigh, NC 27695-8202, USA \\
              \email{ji@ncsu.edu}
}

\date{Received: date / Accepted: date}

\maketitle

\begin{abstract}
We study the light-front covariance of a vector-meson decay constant using a manifestly covariant fermion field theory model in $(3+1)$ dimensions. The light-front zero-mode issues are analyzed in terms of polarization vectors and method of identifying the zero-mode operator
and of obtaining the light-front covariant decay constant is discussed.
%Insert your abstract here. Include keywords, PACS and mathematical
%subject classification numbers as needed.
\end{abstract}

\section{Introduction}
\label{intro}

 Mesonic weak transition form factors and decay constants are
 two of the most important ingredients in studying weak decays of mesons, which
enter in various decay rates. Many theoretical efforts were undertaken to calculate these
observables. The light-front quark model~(LFQM) based on the
LF dynamics~(LFD) has been quite successful in describing various exclusive decays
of mesons~\cite{Jaus99,Cheng04,BCJ_spin1,BCJ_PV,CJ_Bc,CJ_PV,CJ_tensor}.
However, one should also realize that the success of LFD in hadron physics
cannot be realized unless the treacherous points in LFD such as the zero-mode contributions
in the hadron form factors~\cite{Jaus99,Cheng04,BCJ_spin1,BCJ_PV,CJ_Bc,CJ_PV,CJ_tensor}
are well taken care of with proper methods.

In this paper, we study the LF covariance of the vector-meson decay constant using
a manifestly covariant fermion field theory model in $(3+1)$ dimensions.
Although the LF covariant issue for the vector meson decay constant has been raised by
Jaus~\cite{Jaus99} some time ago, systematic analyses
of the zero-modes depending on different polarization vectors
have not yet been explored much. Here, we attempt to
systematically investigate the LF zero-mode issues in terms of polarization
vectors of a vector meson and show
a method of identifying the zero-mode operator to obtain
the LF covariant decay constant
even in the case that there exists a zero-mode contribution.

\section{Manifestly Covariant Calculation}
\label{sec:1}

The decay constant $f_V$ of a vector meson
of mass $M$ and bound state of a quark $q$ of mass $m_1$, and an antiquark ${\bar q}$
of mass $m_2$,
is defined by the matrix element of the vector current

\be\label{eq:1}
\la 0|{\bar q}\gamma^\mu q|V(P,h)\ra
= f_V M \epsilon^\mu(h),
\ee
where the polarization vector $\epsilon$ of a vector meson satisfies the Lorentz condition
$\epsilon \cdot P = 0$.

The matrix element $A^\mu_h \equiv \la 0|{\bar q}\gamma^\mu q|V(P,h)\ra$ is given in
the one-loop approximation as a momentum integral
\be\label{eq:2}
A^\mu_h = N_c
\int\frac{d^4k}{(2\pi)^4} \frac{H_V} {N_p N_k}
{\rm Tr}\left[\gamma^\mu\left(\slash \!\!\!p+m_1 \right)
 \Gamma\cdot\epsilon(h)
 \left(-\slash \!\!\!k + m_2 \right) \right],
\ee
where $N_p  = p^2 -m^2_1 +i\varepsilon$ and
$N_k = k^2 - m^2_2+i\varepsilon$ with $p =P -k$ and $N_c$ denotes the number of colors.
For simplicity in regularizing the covariant loop, we take $n=2$ in a multipole ansatz
$H_V = g/ (N_\Lambda)^n$ for the $q{\bar q}$ bound-state vertex function of a vector meson,
where $N_\Lambda  = p^2 - \Lambda^2 +i\varepsilon$, and $g$ and $\Lambda$ are constant parameters.
The vector meson vertex operator $\Gamma^\mu$ in the trace term
 $S^\mu_h  =  {\rm Tr}\left[\gamma^\mu\left(\slash \!\!\!p+m_1 \right)
 \Gamma\cdot\epsilon(h)\left(-\slash \!\!\!k + m_2 \right) \right]$
is given by
%\be\label{eq:5}
$\Gamma^\mu =\gamma^\mu-(p-k)^\mu/D$.
%\ee
In this work, we shall analyze for the two cases of $\Gamma^\mu$, i.e.,
(1) $\Gamma^\mu=\gamma^\mu$ (i.e. $1/D =0$) case and (2) $\Gamma^\mu$ with constant $D$ factor
(i.e. $D=D_{\rm con}=M + m_1 + m_2$) case
for the explicit comparison between the manifestly covariant calculation and the LF one.
The manifestly covariant result is given by
\be\label{eq:7}
f_V^{\rm Cov} = \frac{N_c g}{4\pi^2 M} \int^1_0 dx\int^{1-x}_0 dy (1-x-y)
\biggl\{ \frac{y(1-y)M^2 + m_1 m_2}{C^2_{\rm cov}}
- \frac{1 + \frac{m_1 + m_2}{D_{\rm con}}}{C_{\rm cov}} \biggr\},
\ee
where $C_{\rm cov} = y(1-y) M^2 - x m^2_1 - y m^2_2 - (1-x-y) \Lambda^2$.

\section{Light-Front Calculation}
\label{sec:2}
Performing the LF calculation in parallel with the manifestly covariant calculation,
we use two different approaches,
i.e. (1) plus component ($\mu=+$) of the currents with the longitudinal
polarization $\epsilon(0)$ and (2) perpendicular components ($\mu=\perp$) of
the currents with the transverse
polarization $\epsilon(\pm)$, to obtain the decay constant. The explicit form of
polarization vectors
of a vector meson is given in~\cite{CJ_PV}.

By the integration over $k^-$ in Eq.~(\ref{eq:2}) and closing the contour in the lower
half of the complex $k^-$ plane, one picks up the residue at $k^-=k^-_{\rm on}$
(on shell value of $k^-$)
in the region $0< k^+ < P^+$ (or $0<x<1$).
Thus, the Cauchy integration formula for the $k^-$ integral in Eq.~(\ref{eq:2}) gives
\be\label{eq:14}
 A^\mu_h = \frac{N_c}{16\pi^3}\int^{1}_0
 \frac{dx}{(1-x)} \int d^2{\bf k}_\perp
 \chi(x,{\bf k}_\perp) S^\mu_h(k^-=k^-_{\rm on}),
\ee
where $S^\mu_h(k^-=k^-_{\rm on})$ is the result of the trace when $k^-=k^-_{\rm on}$ and
%\be\label{eq:15}
$\chi(x,{\bf k}_\perp) = g/[x^3 (M^2 -M^2_0)(M^2 - M^2_{\Lambda})^2]$
%\ee
with $ M^2_{0(\Lambda)} = [{\bf k}^{2}_\perp + m_1^2(\Lambda^2)]/x
 + [{\bf k}^{2}_\perp + m^2_2]/(1-x)$.

%\subsection{$\mu=+$ with longitudinal polarization}

Firstly, using $\mu=+$ with the longitudinal polarization vector $\epsilon(0)$ in Eq.~(\ref{eq:14}),
the decay constant is obtained from the relation
\be\label{eq:17}
f^{(h=0)}_V = \frac{A^+_{h=0}}{\epsilon^+(0)M}.
\ee
For the purpose of analyzing zero-mode contribution to the decay constant,
we denote the decay constant as $[f^{(h=0)}_V]_{\rm val}$
when the matrix element $A^+_{h=0}$ is obtained for $k^-=k^-_{\rm on}$ in the
region of $0<x<1$. Comparing $[f^{(h=0)}_V]_{\rm val}$ with the manifestly
covariant result $f^{\rm Cov}_V$, we find that $[f^{(h=0)}_V]_{\rm val}$  is
exactly the same as $f^{\rm Cov}_V$ when $\Gamma^\mu=\gamma^\mu$ (or $1/D=0$) is used.
The same observation has also been made in~\cite{BCJ_PV}.
However, $f^{(h=0)}_V$ is different from $f^{\rm Cov}_V$ when $D=D_{\rm con}$ is used.
The difference between the two results, i.e.
$f^{\rm Cov}_V - [f^{(h=0)}_V]_{\rm val}$,
corresponds to the zero-mode contribution $[f^{(h=0)}_V]_{\rm Z.M.}$ to the full solution
$[f^{(h=0)}_V]_{\rm Full}=[f^{(h=0)}_V]_{\rm val} + [f^{(h=0)}_V]_{\rm Z.M.}$.

As in the case of zero-mode contribution to the weak transition form factors for
semileptonic $P\to P$ and $P\to V$ decays~\cite{CJ_Bc,CJ_PV},
the zero-mode contribution to $f^{(h=0)}_V$ comes (if exists) from the singular
$p^-$ (or equivalently $1/x$) term in $S^+_{h=0}$ in the limit of $x\to 0$ when $p^- =p^-_{\rm on}$.
For the case of $D=D_{\rm con}$,  we find the following
singular term in $S^{+}_{h=0}$ as follows
\be \label{eq:18}
\lim_{x\to 0}S^{+}_{h=0}(p^- =p^-_{\rm on})
= 4m_1\frac{\epsilon^+(0) p^-}{D_{\rm con}}.
\ee
As we presented in the weak transition form factor calculation~\cite{CJ_Bc,CJ_PV},
we identify the zero-mode operator $[S^{+}_{h=0}]_{\rm Z.M.}$
by replacing $p^-$ with the operator $-Z_2$~\cite{Jaus99,CJ_Bc,CJ_PV}: i.e.
%%
%\be\label{eq:19}
$[S^{+}_{h=0}]_{\rm Z.M.} = 4m_1\frac{\epsilon^+(0)(-Z_2)}{D_{\rm con}}$,
%\ee
where
% \be\label{eq:20}
$Z_2 = x(M^2 - M^2_0) + m^2_1 - m^2_2 + (1-2x)M^2$.
% \ee
%%CR
%That is, the
The zero-mode contribution to the matrix element $A^+_{h=0}$ is given by
\be\label{eq:zm}
 [A^+_{h=0}]_{\rm Z.M.} = \frac{N_c}{16\pi^3}\int^{1}_0
 \frac{dx}{(1-x)} \int d^2{\bf k}_\perp
 \chi(x,{\bf k}_\perp) [S^+_{h=0}]_{\rm Z.M.},
\ee
and the corresponding zero-mode contribution to the decay constant is obtained as
$[f^{(h=0)}_V]_{\rm Z.M.} = [A^+_{h=0}]_{\rm Z.M.}/(\epsilon^+(0)M)$.
Finally, we obtain the full result of the decay constant for the longitudinal polarization as
\bea\label{eq:21}
 [f^{(h=0)}_V]_{\rm Full} &=& \frac{N_c}{4 M \pi^3}\int^{1}_0
 \frac{dx}{(1-x)} \int d^2{\bf k}_\perp
 \chi(x,{\bf k}_\perp)
  \biggl\{ x(1-x)M^2 + {\bf k}^2_\perp
\nonumber\\
&&
 + m_1 m_2
 + (m_1 + m_2)
 \frac{x \left[ {\bf k}^2_\perp + m^2_2 - (1-x)^2 M^2 \right]}{(1-x)D_{\rm con}}
 \biggr\}.
\eea
It can be checked that
Eq.~(\ref{eq:21}) is identical to the manifestly covariant result of
Eq.~(\ref{eq:7}).

Secondly, using $\mu=\perp$ with the transverse polarization vector $\epsilon(+)$,
the decay constant is obtained from the relation
\be\label{eq:22}
f^{(h=1)}_V = \frac{A^\perp_{h=1}\cdot\epsilon^*_{\perp}(+)}{M}.
\ee
In this case, the decay constant $f^{(h=1)}_V$ receives the zero mode from the simple vertex
$\gamma^\perp$ term but not from the term including the $D$ factor.
Following the same procedure as for the case of $f^{(h=0)}_V$, we find the following
singular term in $S^{\perp}_{h=1}$ as
%follows
\be \label{eq:23}
\lim_{x\to 0}S^{\perp}_{h=1}(p^-=p^-_{\rm on})
= 2p^-\epsilon_\perp(+),
\ee
and thus the corresponding zero-mode operator is given by
$[S^{\perp}_{h=1}]_{\rm Z.M.}= 2(-Z_2)\epsilon_\perp(+)$.
Finally, we obtain the full result of the decay constant as follows
\bea\label{eq:25}
 [f^{(h=1)}_V]_{\rm Full} &=& \frac{N_c}{4 M \pi^3}\int^{1}_0
 \frac{dx}{(1-x)} \int d^2{\bf k}_\perp
 \chi(x,{\bf k}_\perp)
 \nonumber\\
 &&\times
 \biggl\{ x M^2_0 - m_1(m_1-m_2) -{\bf k}^2_\perp
  + \frac{ (m_1 + m_2)}{D}{\bf k}^2_\perp
 \biggr\}.
\eea
From Eq.~(\ref{eq:25}) one can check that
that $[f^{(h=1)}_V]_{\rm Full}$ is the same as $[f^{(h=0)}_V]_{\rm Full}$ [Eq.~(\ref{eq:21})],
which confirm the result obtained by Jaus~\cite{Jaus99}.

\section{Conclusion}
In this work, we investigate the LF zero-mode issue for the vector meson decay constant
using two different polarization vectors of a vector meson.
We find that the decay constant obtained from transverse polarization vectors cannot
avoid the zero-mode even at the level of model-independent simple vector meson vertex, i.e.
$\Gamma^\mu=\gamma^\mu$.
Although the decay constant obtained from longitudinal polarization vector may receive a
zero-mode depending on a model-dependent form of $D$ factor,
it is immune to the zero-mode at the level of simple vertex.
The independence of the decay constant on the polarization vectors is also explicitly shown.
%%CR
Our results do not depend on the value of $n (\geq 2)$ in the multipole ansatz and
%Our findings
may give an important guidance on a more realistic model building.

\begin{acknowledgements}
This work  was supported in part by the Korea
Research Foundation Grant(KRF-2010-0009019) and in part
by Kyungpook National University Research Fund, 2012
\end{acknowledgements}



\end{document}